\documentclass[aps,prd]{revtex4}

\usepackage{graphicx}

\begin{document}

\def\be{\begin{equation}}
\def\fe{\end{equation}}
\def\bea{\begin{eqnarray}}
\def\fea{\end{eqnarray}}

\title{How viscous is a superfluid neutron star core?}

\author{N.~Andersson$^*$, G.~L.~Comer$^\dag$ and K. Glampedakis$^*$}
\affiliation{$^*$School of Mathematics, University of 
Southampton, Southampton SO17 1BJ, UK \\ 
$^\dag$Department of Physics, Saint Louis University, St.~Louis 
MO 63156-0907, USA}

\begin{abstract}
We discuss the effects of superfluidity on
the shear viscosity in a neutron star core. Our 
study combines existing theoretical results for the viscosity coefficients
with data for the various superfluid energy gaps into a consistent description.
In particular, 
we  provide a simple model for the electron viscosity
which is relevant both when the protons form a normal fluid and
when they become superconducting. This model explains in a clear way
why proton superconductivity leads to a 
significant strengthening of the shear viscosity.
We present our results in a form which permits the use of 
data for any given modern equation of state
(our final formulas are explicitly 
dependent on the proton fraction). We discuss a simple 
description of the relevant superfluid pairing gaps, and construct a 
number of models (spanning the range of current uncertainty)
which are then used to discuss the superfluid suppression of shear
viscosity. We conclude by a discussion of
a number of  challenges that must be met
if we are to make further progress in this area of research.
\end{abstract}

\date{\today}

\maketitle

\section{Introduction}

Neutron stars are often thought of as exciting cosmic laboratories.
This is natural since their description depends on much 
complex physics \cite{latt}: With a mass of about one and a half times that of the Sun 
compressed inside a radius of ten kilometers or so, they are compact enough to 
require a fully general relativistic description. With central densities several times the 
nuclear saturation density it may be energetically favourable for exotic 
phases of matter, like kaon condensates, hyperons and/or deconfined quarks, to be 
present in the core. With temperatures much below the Fermi temperature for the various 
constituents, neutron stars are cold on the nuclear scale which means that 
the presence of both solid (the outer layers form a kilometer sized nuclear lattice)
and superfluid regions is expected. Neutron star dynamics is also intriguing.
The obvious example of this is the glitches, which are taken as evidence
of at least two weakly coupled interior components, observed in a number of
radio pulsars \cite{lyne}. In addition, the evolution of the sample of the neutron star population
which is spinning rapidly may be affected by both
hydrodynamical and/or radiation-driven instabilities \cite{narev}.
This is particularly interesting from the point of view of gravitational-wave
observations. With a generation of large-scale gravitational-wave interferometers
now reaching design sensitivity, it is worth emphasising that 
a signal from an (unstable?) neutron star pulsation mode
could provide a unique, and perhaps the only, probe of 
the bulk motion of matter inside such stars. 
The information that could be gleaned from such data would be truly unprecedented \cite{astero,acprl}.

As a useful illustration of a problem which plays a key role in neutron star dynamics, 
let us consider the coupling between the elastic crust of 
nuclei and the fluid core. 
The relevant coupling timescale impacts on, for example, the damping rate of 
oscillations driven unstable by gravitational-radiation emission 
(eg. the r-modes \cite{narev}), 
the relaxation following a pulsar glitch
\cite{easson,cheng,alp}, and possible neutron star free precession \cite{ianj,link,levin}.
In its simplest form the problem is analogous to the classic 
``spin-up problem'' in fluid dynamics, where one studies
the rotation rate of a viscous
fluid following a change in  rotation of the container. 
It is well-known that the fluid velocity changes due to the formation of a 
so-called Ekman layer at the interface. 
The role of this viscous boundary layer is to 
ensure that the fluid motion satisfies a no-slip condition at the interface. 
Its presence results in a coupling to the bulk of the fluid on a timescale
\be
t_\mathrm{E} \sim t_\nu \left( {\delta_\mathrm{E} \over R} \right) \qquad \mbox{ with } \qquad \delta_\mathrm{E} = 
\left( { \eta \over \rho \Omega} \right)^{1/2} 
\fe
where $\delta_\mathrm{E}$ is the width of the Ekman layer~\footnote{It is worth pointing out that one would not expect an Ekman layer to form 
on the crust-side of the phase transition. The main reason for this is that the elastic shear stresses will overwhelm
those associated with the viscosity. Hence,  all quantities in the equation should be evaluated for the core fluid at the base of the crust.},  
$R$ is the 
radius of the container, $\Omega$ is the rotation rate, $\rho$ is the mass density at the interface and $\eta $ is the shear viscosity coefficient.
For a neutron star, where $\delta_\mathrm{E}/R\sim 10^{-6}$ it is clear that the resulting timescale is much shorter than 
that of viscous diffusion, i.e. 
\be
t_\nu \sim { R^2 \rho \over \eta} 
\fe 
Analogously, in the case of an oscillation in the fluid 
the Ekman layer provides an efficient damping mechanism, see \cite{narev} for a recent review.
 
However,  it is a serious 
oversimplification to model the core-crust interface as a 
viscous fluid/solid wall transition. 
In particular, the fact that the 
crust is expected to be permeated by superfluid neutrons
should also be considered. Similarly, the multi-fluid nature of the
expected superfluid neutron/superconducting proton mixture in the 
outer core must be accounted for. This involves many difficult
issues associated with the presence of rotational vortices in the 
neutron superfluid, and magnetic fluxtubes in the proton 
fluid (if it forms a type-II superconductor). The interactions involving the 
vortices, leading to the non-dissipative entrainment effect \cite{cj} as well as
the mutual friction \cite{men91}, and the potential pinning of vortices to the nuclear lattice
\cite{cheng,alp}
are not yet understood in detail.    
It is clear that one faces 
a problem which stretches our understanding of neutron 
star physics considerably. In a series of papers, we aim to make progress by 
considering each of the relevant issues in turn. 

The purpose of the present paper is to discuss the shear viscosity in a superfluid neutron star.
In doing this, we are taking a small step towards a fully consistent 
description of dissipative neutron star cores. As a first step in this
development we revisit the standard shear viscosity (ignoring for the moment 
the multi-fluid dynamics
aspects), and ask how well 
understood the relevant viscosity coefficients are. This leads to
a set of useful formulas that can readily be used in future work.
In addition, our discussion identifies a number of challenging 
research problems that require further attention.   

\section{Revisiting the shear viscosity}

We begin by making  the following observation:
In their by now classic paper on the effect of viscosity on 
damping of oscillations in superfluid neutron stars, Cutler 
and Lindblom~\cite{cl87} state:
``It is interesting to note that contrary to our experience 
with other superfluids like ${\rm He}_{\rm 4}$, neutron star 
matter becomes more viscous in the superfluid state than it 
was in the normal state."
This, at first sight, counter-intuitive result provides ample motivation for our investigation.

 Estimates of the effect that 
shear viscosity has on the fluid motion in neutron stars have so far
almost exclusively been 
based on the  work of Flowers and Itoh \cite{fi76,fi79}, who
calculated the relevant transport coefficients at supranuclear
densities. Based on the results of these papers, Cutler and Lindblom \cite{cl87}
discussed the viscous damping  of oscillations in the neutron star fluid. 
Their analysis was based on, first of all, a fit to the result in \cite{fi79} 
for the total shear viscosity in normal fluid neutron stars, which can be written
\begin{equation}
\eta_n \approx 2\times 10^{20} \rho_{15}^{9/4} T_8^{-2} \ \mbox{g/cm s}
\label{eta_n}\end{equation}
where $\rho_{15}= \rho/10^{15}\ \mbox{g/cm}^3$ and $T_8 = T/10^8$~K.  
As stated by Flowers and Itoh (and as is also evident from Figure~2 in 
\cite{fi79}),  the dominant contribution to $\eta_n$ is due to neutron-neutron scattering. 
This result makes sense since the neutrons make up the bulk of the fluid at 
the relevant densities (and more exotic massive particles like hyperons are not being considered). 
However, the neutrons in the outer core are expected to become superfluid as soon as the star cools 
below (say) $10^9$~K, i.e. soon after its birth in a supernova
core collapse. Below the neutron superfluid transition temperature the neutron-neutron scattering is suppressed and the dominant contribution to the 
shear viscosity is made by  scattering processes involving
the  relativistic electrons. 
Cutler and Lindblom \cite{cl87} argue that the relevant estimate 
(for electron-electron scattering) in a superfluid 
neutron star is
\begin{equation}
\eta_{ee}^{\rm CL} \approx 6\times 10^{20} \rho_{15}^{2} T_8^{-2} \ \mbox{g/cm s}
\label{eta_CL}\end{equation}
This estimate for $\eta_{ee}$ has since been used in a number of contexts. 
The puzzling fact that $\eta_{ee}^{\rm CL}>\eta_n$ 
throughout a typical neutron star 
does not seem to have caused much concern.
In addition to possible confusion about this result, the estimate (\ref{eta_CL})
is not quite 
satisfactory. One reason is that it is associated with a particular 
supranuclear equation of state (derived by Baym, Bethe and Pethick in 1971 \cite{bbp}).
In practise, one would like to 
have an estimate which explicity includes the dependency on the electron (=proton) fraction. 
This is important since one may then be able to better understand the 
effect of varying the supranuclear equation of state (and the proton fraction) 
within the current range of uncertainty. 

To derive the relevant estimate, we 
take as our starting point the general formula 
(cf.  \cite{fi76,EP79,NP84})
\be
\eta_e = { n_e p_e^2 \over 5 m_e^\ast } \tau
\label{etaformula}\fe
where $n_e$ is the electron number density, 
$p_e = \hbar k_F = \hbar (3\pi^2 n_e)^{1/3}$ is the corresponding 
Fermi momentum, $m_e^\ast = p_e/c$ is the 
effective electron mass and $\tau$ is the timescale for momentum transfer due to the 
various scattering processes in which the electrons are involved.
In a normal fluid neutron star core, scattering off of the (nonrelativistic) 
protons provides the main channel for momentum exchange. However, if the 
protons become superconducting then
electron-electron scattering takes over as the dominant contribution to 
$\tau$ \cite{easson,fi79}. 

In order to model the resultant shear viscosity, we use the 
timescale for electron-proton scattering derived by Easson and Pethick
\cite{EP79}:
\be
\tau_{ep} = { 4 \over \pi^2 \alpha^2} \left( { \epsilon_{Fp} \over k_B T} 
\right)^2 { k_{ft} \over c k_F^2} 
\fe
where $\alpha = 1/137$ is the fine-structure constant, $\epsilon_{Fp}$ 
is the proton Fermi energy and  $k_{ft}$ is the Thomas-Fermi screening 
wave-vector (to be discussed later). Note that we are assuming that the matter is charge neutral, i.e. we take $n_e=n_p$.

We can extract an analogous formula for electron-electron scattering from 
Eq.~(91) in \cite{fi76}. This equation can be written 
\be
{ 1 \over \eta_{ee}} = { 15\over 2} { \pi^4 \alpha^2 \over \hbar k_F^3} \left( { k_B T \over p_e c }\right)^2 
\left( {2 k_F\over k_{ft}} \right) \left[ { 5\over 2} + 3 \left( 
{m_e c\over p_e} \right)^2 + \left( 
{m_e c \over p_e} \right)^4 \right]
\label{shear}\fe
Noting that 
\be
{ m_e c \over p_e} \approx 8\times 10^{-4} \left( { n_e \over 1 \mbox{ fm}^{-3} } \right)^{-1/3}
\approx 4.5 \times 10^{-3} \left( {x_p \over 10^{-2}} \right)^{-1/3} \rho_{15}^{-1/3}
\fe
where $x_p$ is the proton (electron) fraction, we see that we 
only need to retain the 
leading order contribution (in $m_e c/p_e$).  By comparing 
(\ref{etaformula}) and (\ref{shear})   we then find that
\be
\tau_{ee} = { 4 \over 10 \pi^2 \alpha^2} \left( { \epsilon_{Fe} \over k_B T} 
\right)^2 { k_{ft} \over c k_F^2} 
\fe
where $\epsilon_{Fe} = \hbar c k_F$ for relativistic electrons.

In order to combine the two results for $\tau_{ep}$ and $\tau_{ee}$, we need to note that individual 
scattering processes add like ``parallel resistors'' 
, cf. \cite{fi76}, which means that we have
\be
\tau = \left[ { 1 \over \tau_{ee} } + {  1 \over \tau_{ep} }\right]^{-1}
\label{tautot}\fe
This  shows that the most important contribution to the shear viscosity
comes from the most frequent scattering process (as one would intuitively expect). Using 
\be
{ \tau_{ee} \over \tau_{ep} } = { 1 \over 10} { \epsilon_{Fe} \over 
\epsilon_{Fp} } =  { 1 \over 10} { 2m_p^\ast c \over \hbar k_F } 
\fe
where   $\epsilon_{Fp} = \hbar^2 k_F^2/2m_p^\ast$ and $m_p^\ast$ is the effective proton mass, 
we find that $\tau_{ee} >> \tau_{ep} $ under normal circumstances. 
Thus the electron-proton scattering provides the dominant contribution to
the electron shear viscosity, in accordance with the discussion of Flowers and Itoh 
\cite{fi79}. 

We now want to account for the likely possibility that the protons are 
superconducting. In that case the electron-proton scattering will be 
suppressed, essentially because there will be fewer states available for the
protons to scatter into. In order to allow for the transition to 
proton superconductivity, we introduce a suppression factor ${\cal R}_p$ such that
\be
\tau_{ep} \longrightarrow {\tau_{ep} \over {\cal R}_p} 
\fe
Far below the critical transition temperature at which the protons become 
superconducting we should have ${\cal R}_p\to 0$, and we see from  (\ref{tautot}) that the 
electron-electron scattering then dominates the shear viscosity \cite{easson}.

Combining the above results we obtain the following final formula
\be
\tau = { 4 \over 10 \pi^2 \alpha^2} \left( { \epsilon_{Fe} \over k_B T} 
\right)^2 { k_{ft} \over c k_F^2} \left[ 1  + { \mathcal{R}_p \over 10} 
 \left( { \epsilon_{Fe} \over 
\epsilon_{Fp} }\right)^2 \right]^{-1}
\label{tau_eta}\fe
which should be used together with (\ref{etaformula}).
It should be noted that we have neglected the contribution from electron-neutron scattering. This should always be a valid approximation. 

Let us now turn to the screening factor $k_{ft}$. Electron scattering is screened by both electrons and 
protons, and the
relevant screening wave vector is given by~\footnote{In a draft version of this paper we followed 
Gnedin and Yakovlev \cite{gnedin} who argued that the protons should not contribute to the screening once they become superconducting. 
We are grateful to the authors of \cite{gnedin} for pointing out to us that this result was erroneous, and for helpful discussions
concerning the screening in general.}
\be
k_{ft}^2 = { 4\alpha \over \pi } k_F^2 \left( 1 + { m_p^\ast c \over \hbar k_F} \right)
\fe 
The first term in the bracket is due to the electron screening, 
while the second is due to the protons. It is straightforward to show that the latter
tends to dominate under the conditions that prevail in a neutron star core.

\begin{figure}[h]
\centerline{\includegraphics[height=8cm,clip] {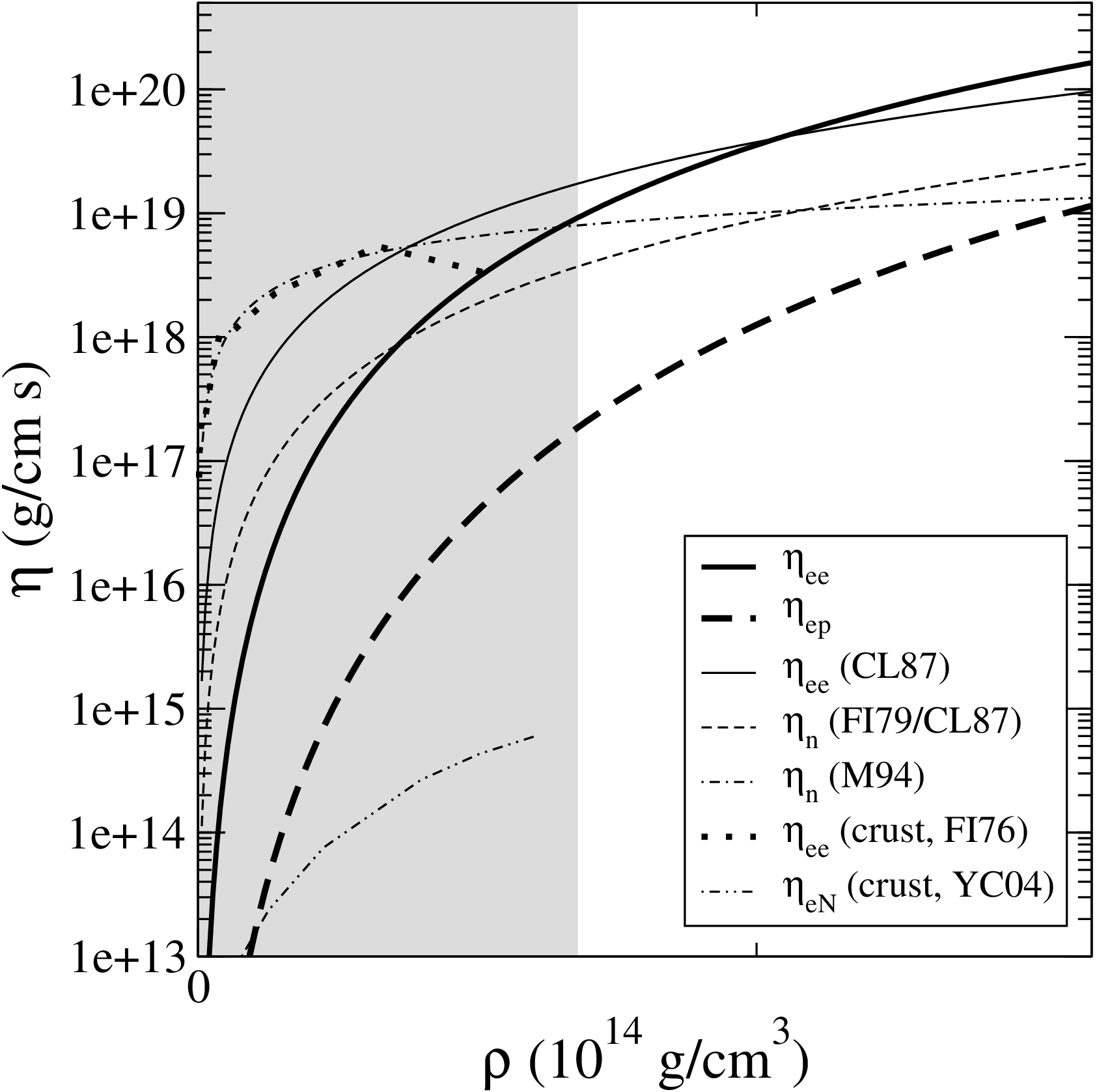} }
\caption{An illustration of the various shear viscosity estimates for a temperature of $10^8$~K. 
The thick solid line shows 
our estimate $\eta_{ee}$ while the thick dashed curve corresponds to 
$\eta_{ep}$ (where we have taken $m_p^\ast = m_p$ in evaluating the screening factor). 
These estimates are compared to i) the result of Cutler and Lindblom \cite{cl87} for 
electron-electron viscosity (CL87, thin solid line), ii) the total shear viscosity in a normal
fluid neutron star, which is dominated by neutron-neutron scattering \cite{fi79,cl87} (FI79/CL87, thin dashed line), 
iii) an alternative calculation of the neutron shear viscosity due to Mornas \cite{mornas} (M94, dot-dashed thin line), 
iv) the results obtained by Flowers and Itoh \cite{fi76} for the electron-electron shear viscosity in the crust
region (crust, FI76, thick dotted curve) and v) a recent estimate by Yakovlev and Chugunov \cite{yc04} which accounts for the fact that
in the crust the 
electrons mainly scatter off of nuclei in the crust (YC04, double 
dot-dashed line). The crust region is indicated by the grey region, and we have 
assumed that the core-crust transition takes place at a density of $1.7 \times 10^{14} \mbox{ g/cm}^3$. 
All estimates (apart from the crust results) were determined using the simple PAL equation of state model discussed
in the text.
It is  interesting to note that the 
Flowers and Itoh crust result dips to join our estimated $\eta_{ee}$ as the base of the crust is approached.
It is also worth noting the several order of magnitude discontinuity between the dominant 
viscosity contributions at the core-crust interface. Comparing the result of Yakovlev and Chugunov to the other estimates, we see that 
there will be a step of more than two orders of magnitude even when the protons are normal and provide the main scattering agent
for electrons in the core.  }
\label{ee-fig}\end{figure}

Having introduced  the key ingredients, we can discuss two 
limiting cases. In the case of normal protons we take ${\cal R}_p = 1$
in (\ref{tau_eta}). This leads to 
\be
\eta_{ep}  \approx 1.8 \times 10^{18} \left( { x_p \over 0.01 } \right)^{13/6} 
\rho_{15}^{13/6}  T_8^{-2} \mbox{ g/cm s}
\label{eta_est}
\fe
Here, and in the following, we have taken (for simplicity) $m_p^\ast = m_p$ in evaluating $k_{ft}$.
Our estimate (\ref{eta_est}) agrees with the result derived by Easson and Pethick \cite{EP79}. 
In the opposite limit the protons are (strongly) superconducting, which means that 
${\cal R}_p = 0$, and we arrive at the estimate
\be
\eta_{ee} \approx 4.4 \times 10^{19} \left( { x_p \over 0.01 } \right)^{3/2} 
\rho_{15}^{3/2}  T_8^{-2} \mbox{ g/cm s}
\label{eta_ee}\fe
This result should be compared to the formula used by Cutler and Lindblom 
\cite{cl87} in 
deriving (\ref{eta_CL}). One can easily show that the two results are compatible. 
This means that the differences that can be seen in Figure~\ref{ee-fig}, where we
compare the various  coefficients, is entirely due to the fact that different equations of state
(eg. proton fractions) are used. 
This figure shows clearly that the (potentially 
puzzling) result that the electron-electron viscosity in the neutron star core
is stronger than that due to 
normal neutron-neutron scattering remains true.  
We also learn that the key reason for the emergence of the dominant electron
shear viscosity is the superconductivity of the protons. 

The results illustrated in Figure~\ref{ee-fig} (and others discussed
throughout the paper)
were obtained using one of the phenomenological 
PAL equations of state (with compression modulus $K=240$ MeV) \cite{PAL}. In order to obtain a suitably simple 
model we have followed Kaminker et al \cite{KHY01} 
who provide the following fit for the 
two nucleon number densities;
\be
n_x = a \rho_{14}^b/ ( 1 + c \rho_{14} + d \rho_{14}^2)  
\label{palfit}\fe
where
$$
a = 0.1675 \ , \qquad b = 1.8185 \ , \qquad c = 2.0288 \ , \qquad d = 0.02444 \qquad \mbox{for} \ x=n
$$
and
$$
a = 0.0006823 \ , \qquad b = 2.6727 \ , \qquad c = 0.1946 \ , \qquad d = 0.01604 \qquad \mbox{for} \ x=p
$$
The total number density $n_b$ and the proton fraction $x_p$ follow
immediately from $n_b = n_n + n_p$ and $x_p = n_p/n_b$, respectively.

In one of the few alternative calculations of the neutron shear viscosity
(that we are aware of), Mornas \cite{mornas} derives the 
following expression [cf. her Eq.~(4.9b)]
\be
\eta_n \approx 2.4\times 10^{19} \rho_{15}^{3/5} T_8^{-2}  \mbox{ g/cm s}
\label{morn}\fe 
It is notable that the scaling with the density is very different 
from that in Eq.~(\ref{eta_n}), possibly because of the density dependence 
of the effective neutron mass. As can be seen from Figure~\ref{ee-fig}, Eq.~(\ref{morn})
leads to a considerably weaker viscosity at high density. It should also be noted that
Mornas's formula leads to results that are very close to the 
electron-electron scattering results for the crust region 
determined by Flowers and Itoh \cite{fi76}. We 
have no rational explanation for why this should be the case
(in fact, it must be a coincidence), but it is a 
potentially useful observation.  

\section{Modelling the effects of superfluidity}

The formulas given in the previous section
provide a useful step towards
modelling the dynamics of realistic neutron stars since they
allow us to determine a consistent electron viscosity for various supranuclear
equations of state and core temperatures. 
Of course, in order to use these formulas we need not only 
the total number density and the proton fraction, we also 
need to i) determine whether the protons and/or neutrons are superfluid, and  
ii) if so, quantify the associated suppression factors.
This forces us to venture into the thorny area of nucleon superfluidity 
(see \cite{LS00} for a recent review).

In this section, we summarise the current understanding of the
relevant superfluid energy gaps and discuss 
the suppression factors we require to 
complete our model.  
From a survey of the vast superfluid gap-literature 
\cite{CCY72,AO85a,AO85b,CCKS86,BCLL90,BCLL92,CCDK93,WAP93,KKC96,EEHO96a,EEHO96b,SCLBL98,EEHO96c,EH98,BEEHS98}
it is clear that the determination of the relevant parameters constitute a 
serious challenge.  In many ways this is
not too surprising. After all, significant uncertainties 
concerning the  supranuclear equation of state remain. 
Given the fact that the 
superfluid energy gaps are  sensitive to the internal constitution of
the star, we have to accept a range of possibilities at the present time. 

It is natural to adopt a strategy similar to that used in the (closely
related) discussion of neutron star cooling, see for example the work 
by Kaminker and colleagues \cite{KHY01,YKG01,KYG02,YGKP02} and 
the recent discussion of a ``minimal cooling paradigm'' by Page et al \cite{page}. That is, 
to consider the plausible range of superfluid parameters suggested by the 
theoretical studies and ask whether observations provide useful 
constraints on the models. Since our focus is on the shear viscosity in a 
neutron star core, the relevant observations concern pulsar 
glitches and potential 
future data for core fluid oscillations. 
In particular, it is known that
gravitational waves from internal dynamics could provide an excellent 
probe of the core physics \cite{acprl}. 


In principle, superfluidity in the singlet state
appears if the formation of Cooper pairs from 
particles with opposite momenta and spins leads to a lowering 
of the ground state energy. A
key parameter in any discussion of nucleon superfluidity is the 
energy gap $\Delta (k)$, which corresponds to the 
energy needed to create a quasiparticle of momentum $k$
in the superfluid \cite{CCDK93}. 
The gap on the Fermi surface $\Delta(k_F)$ can be 
interpreted as half the energy required to break the
Cooper pairs. In the following we will focus on estimates of
this quantity since it allows us to approximate both the 
critical temperature at which the matter becomes superfluid
and the reduction factors
required in our analysis of shear viscosity.

The strong interaction provides several channels in which pairing is possible. 
The indications are that Cooper pairs with zero angular momentum 
in a spin-singlet state $^1S_0$ should form at low momenta. However, 
since the effective  interaction
is expected to become repulsive near the nuclear saturation density,
the corresponding energy gap will disappear as the density increases \cite{KKC96}. 
As a result, because of the comparatively low proton number density one would 
expect proton $^1S_0$ pairing at 
densities too high for neutron pairing in that phase \cite{AO85a,AO85b}. 
All existing 
models suggest that the dominant proton pairing takes 
place well inside the core-crust transition. 
At the same time, the 
energetic advantage of neutron pairing in the anisotropic
$^3P_2$ state at higher densities may be attributed to 
the tensor force \cite{CCKS86}.

Given the  practical difficulties associated with numerical
many-body calculations, 
most studies have been based on the ``standard'' BCS theory, including 
more and 
more refined pairing interactions by adding successive corrections to the 
bare nucleon-nucleon potential \cite{BCLL90}. The next order
beyond the bare interaction in this scheme corresponds to 
various medium effects.  These are often accounted for by renormalising 
the single-particle energies, eg. by lowering the effective nucleon mass. 
It is by now well-known that this leads to a
smaller energy gap. The key factor determining the 
effective mass is the number of
interaction partners of opposite type that a given
nucleon has, i.e. the number of neutrons per proton and vice versa. 
The greater the number, 
the greater
will be the  effect of the medium in reducing the effective mass.
Hence, the effective 
proton mass tends to be smaller than that for neutrons, 
and one may expect the maximum gap for the  $^1S_0$ proton 
superfluid to be smaller than that for 
neutrons. Similarly, a major difference between results for pure 
neutron matter and nuclear matter is due to the induced lowering of the effective
neutron mass by the relatively small proton contaminant \cite{CCKS86}.

While the results for the bare interaction have converged towards 
a predicted maximum $^1S_0$ neutron 
gap of about 3~MeV at $k_F\approx 0.85 \mbox{ fm}^{-1}$ \cite{KKC96,EEHO96b,SCLBL98,EH98}
(model $A$ in Table~\ref{gaptable}), the higher order calculations
still lead to a range of possibilities. Particularly important is the fact that 
pairing polarises the medium \cite{CCY72}. After accounting 
for the polarisation effects one finds that
the maximum gap has been quenched to nearly 1 MeV at 
$k_F \approx 0.7-0.8 \mbox{ fm}^{-1}$ \cite{CCDK93,WAP93,SCLBL98}
(models $a-d$ in Table~\ref{gaptable}). A nice explanation for this 
effect is provided by Lombardo and Schulze \cite{LS00} who show that 
 low density polarisation effects  suppress the BCS gap by 
roughly a factor $(4e)^{-1/3}\approx 0.45$.

The $^1S_0$ proton and $^3P_2$ neutron gaps provide further challenges. In the former 
case one must take into account the background neutron influence (eg. the 
polarisation terms), while in the latter one must solve the anisotropic 
gap equations (in principle ten coupled equations)
in a consistent way. 
In the case of the proton gap the result is particularly sensitive to 
the so-called ``symmetry energy''. This is expected since this parameter
governs the proton abundance, cf. \cite{BCLL92,PAL}. The available results 
suggest a maximum gap for proton $^1S_0$ pairing of approximately 1~MeV 
at $k_F \approx 0.4-0.5 \mbox{ fm}^{-1}$ 
(models $e$ and $g$ in Table~\ref{gaptable}). This is 
roughly a factor of 3 smaller than the maximum neutron gap. 
The difference can be understood from the fact that the 
effective proton mass is smaller than the effective neutron mass. 
(In a pure proton medium the proton gap should, because of the charge symmetry of the nuclear 
interaction, be identical to the neutron gap in a pure neutron fluid \cite{CCY72}.)
One would expect
polarisation effects to further influence the proton gap, 
leading to a 
suppression (at least?) similar to that found for 
the $^1S_0$ neutron channel (a factor of a few) \cite{CCDK93}. 
Thus the maximum
proton gap may be reduced to 
perhaps 0.2--0.3 MeV (a level similar to model $f$ in Table~\ref{gaptable}).

In order to study the 
anisotropic $^3P_2$ neutron gap in detail one must extend BCS theory, and
replace 
the single gap equation for the $^1S_0$ case  by ten coupled
equations. In fact, the added 
attraction from the tensor coupling is essential for the existence
of superfluidity in this state \cite{EEHO96a}. As in the case of
the isotropic gaps, one finds that the $^3P_2$ gap is
reduced significantly (by a factor of two or so) by 
the lower neutron effective mass in nuclear matter.
The available results for the $^3P_2$ gap illustrate 
the extent to which this problem remains to be understood: 
The calculation of Elgar\o y et al \cite{EEHO96a}  
(models $k$ and $l$ in Table~\ref{gaptable}) suggests a much smaller gap 
than that predicted in the work by Baldo and colleagues \cite{BEEHS98}
(models $h-j$ in Table~\ref{gaptable}).
The difficulties associated with superfluid pairing in the $^3P_2$ channel
are exacerbated by the fact that relativistic effects come into play at the 
relevant densities. 
While the $^1S_0$ results remain largely unaffected by the inclusion of 
relativistic effects (model $f$  in Table~\ref{gaptable}), the associated change in the single-particle energies 
reduces the $^3P_2$ gap by about a factor of two
 \cite{EEHO96c} (model $m$ in Table~\ref{gaptable}).
Finally, 
in contrast to the case for the $^1S_0$ gaps, it has been suggested that polarisation effects 
(which have yet to be accounted for) may 
increase the $^3P_2$ neutron gap. 

\begin{table}[h]

\begin{tabular}{ccccccl}
\hline
model & $\Delta_0$ (Mev) & $k_1 \mbox{ (fm}^{-1})$ & $k_2 \mbox{ (fm}^{-1})$ & $k_3\mbox{ (fm}^{-1})$ & $k_4\mbox{ (fm}^{-1})$ & Reference/Comments \\
\hline
$A$	& 9.3 & 0.02 & 0.6 & 1.55 & 0.32 & \cite{KKC96,EEHO96b,SCLBL98,EH98}  (bare interaction)\\
\hline
$a$	& 68  & 0.1  & 4 & 1.7 & 4 & \cite{WAP93} \\
$b$ 	& 4   & 0.4  & 1.5 & 1.65 & 0.05 & \cite{SCLBL98} \\
$c$	& 22  & 0.3 & 0.09 & 1.05 & 4 & \cite{CCKS86} (Reid potential w. $m_\ast$) \\
$d$	& 2.9 & 0.3 & 0.017 & 1.3 & 0.07 & \cite{CCKS86}  (Reid potential w. polarisation) \\
\hline
$e$	& 61 & 0 & 6 & 1.1 & 0.6 & \cite{EEHO96c} (relativistic)\\
$f$	& 55 & 0.15 & 4 & 1.27 & 4 & \cite{AO85a} (OPEG potential w. $m_\ast=0.6$)\\
$g$        & 2.27 & 0.1 & 0.07 & 1.05 & 0.25 & \cite{BCLL92} \\
\hline
$h$	& 4.8 & 1.07 & 1.8 & 3.2 & 2 & \cite{BEEHS98} (Bonn B potential, free spectrum) \\
$i$ 	& 10.2 & 1.09 & 3 & 3.45 & 2.5 & \cite{BEEHS98} (Argonne $V_{14}$ potential, free spectrum) \\
$j$ 	& 2.2 & 1.05 & 1 & 2.82 & 0.6 & \cite{BEEHS98}  (Paris potential, free spectrum) \\
$k$	& 0.425 & 1.1 & 0.5 & 2.7 & 0.5 & \cite{EEHO96a} (non-relativistic)\\
$l$ 	& 0.068 & 1.28 & 0.1 & 2.37 & 0.02 & \cite{EEHO96a} (non-relativistic) \\
$m$ 	& 2.9	& 1.21 & 0.5 & 1.62 & 0.5 &   \cite{EEHO96c} (relativistic)\\
\hline 	
\end{tabular}
\caption{Detailed parameters for our  various gap models constructed from (\ref{deltf}).
The models are based on  calculations in the given references and represent the 
current range of possibilities. Model $A$ is for the bare interaction and is relevant in a 
pure neutron (proton) medium. Models $a-d$ are for the $^1S_0$ neutron pairing, while
models $e-g$ correspond to the $^1S_0$ proton results and models $h-m$ are for the $^3P_2$ neutron 
channel. 
 }
\label{gaptable}
\end{table}

In order to consider various gap-models in our analysis, 
without complicating things excessively, we take a lead from  
Kaminker and colleagues \cite{KHY01,YKG01,KYG02,YGKP02} 
and represent the energy gap (at the Fermi surface) by the 
phenomenological formula
\be
\Delta (k_F) = \Delta_0 { (k_F-k_1)^2 \over (k_F-k_1)^2 + k_2} 
{ (k_F-k_3)^2 \over (k_F-k_3)^2 + k_4}
\label{deltf}
\fe
where $k_F$ is the Fermi momentum of the relevant nucleon. 
This expression makes sense because we know from the weak-coupling 
formula of BCS theory that $\Delta  \sim k_F^2$ at low densities
(leaving out an exponential factor which further suppresses the gap as $k_F\to 0$). 
The results in the literature also indicate that the bare interaction
gap function 
is roughly symmetric \cite{KKC96,EEHO96b,SCLBL98,EH98}. This suggests that $\Delta \sim (k_F-k_3)^2$, where the gap vanishes for $k_F>k_3$, for higher
densities. The additional parameters $k_2$ and $k_4$ permit us to adapt the shape of $\Delta$ 
according to various results which incorporate, for example, polarisation 
effects. 

\begin{figure}
\centerline{\includegraphics[height=6cm,clip] {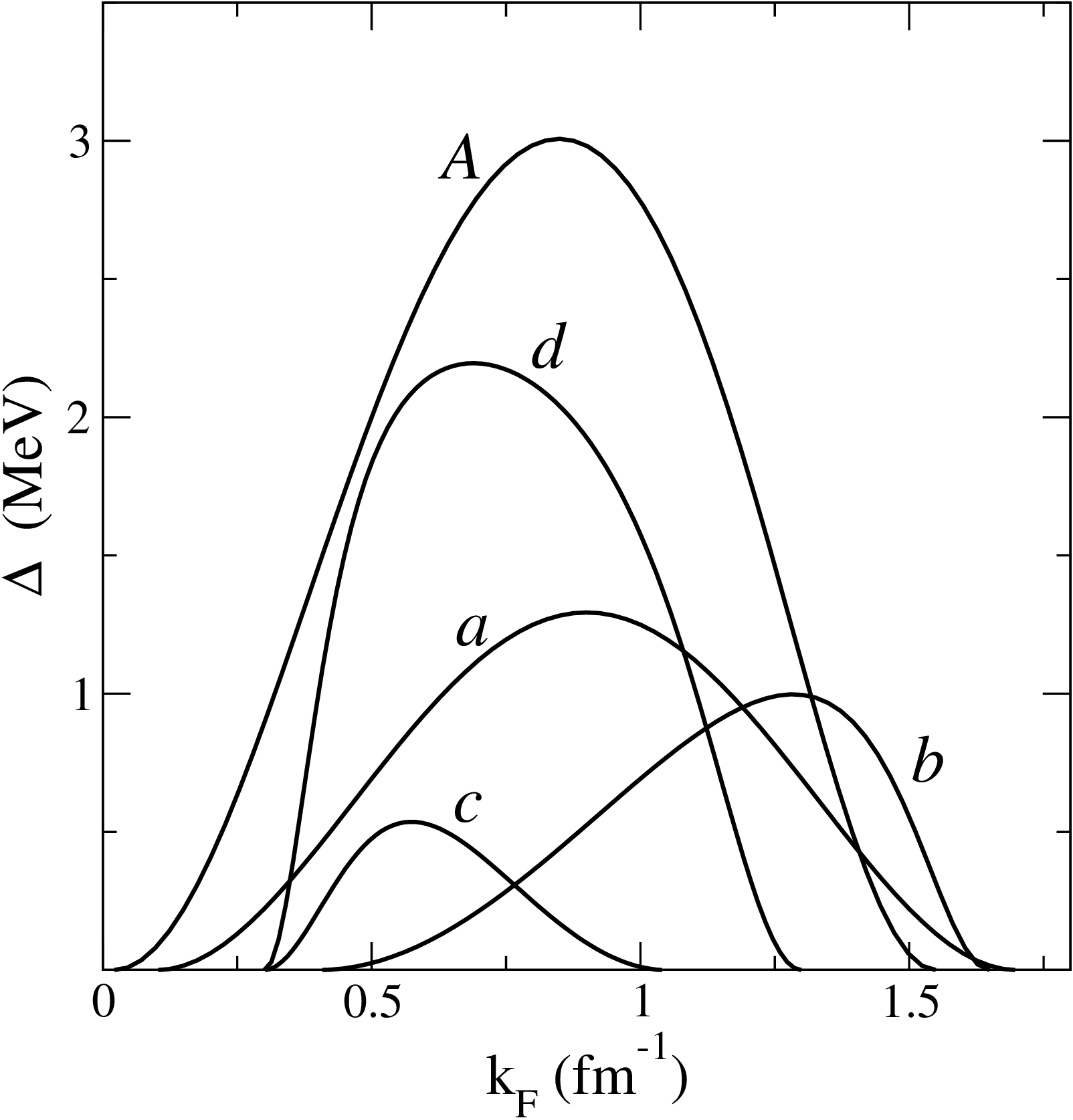} \includegraphics[height=6cm,clip] {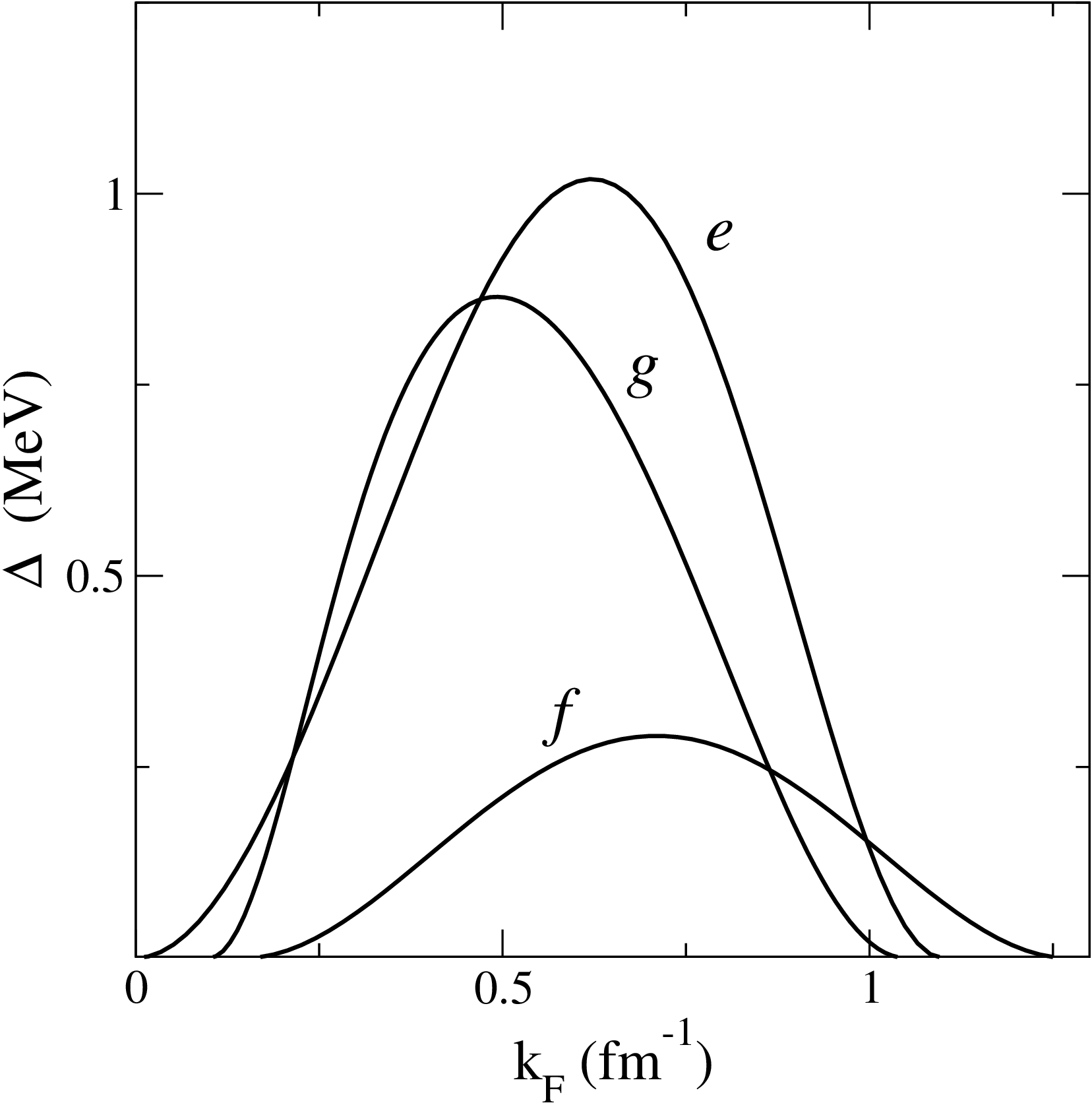}}
\caption{The various energy gaps for the $^1S_0$ pairing, corresponding to the data in Table~\ref{gaptable}, 
are shown as functions of $k_F$. The left panel corresponds to the neutrons, 
while the right panel shows the proton results. }
\label{1s0gaps}
\end{figure}

\begin{figure}
\centerline{\includegraphics[height=6cm,clip] {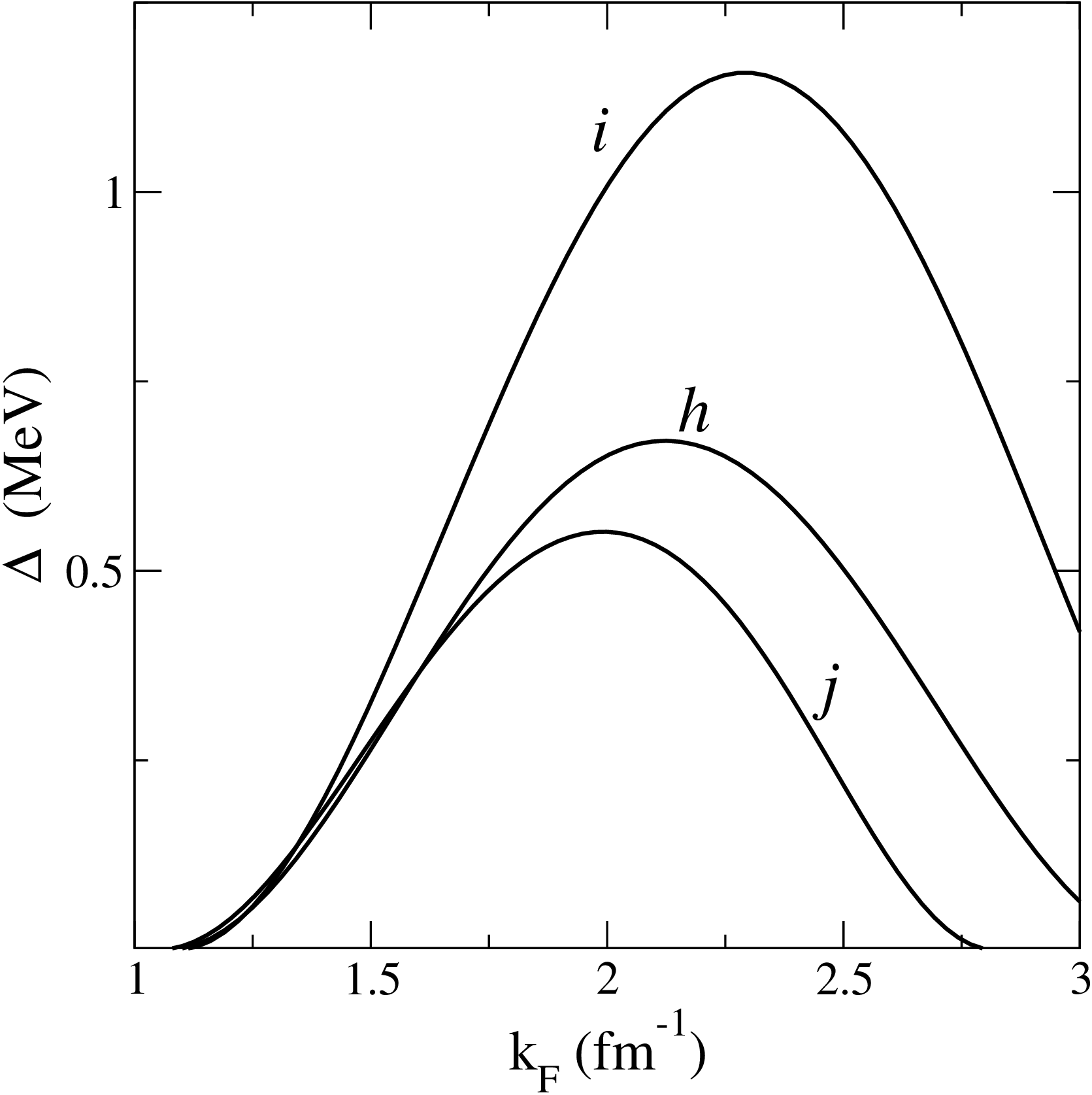} \includegraphics[height=6cm,clip] {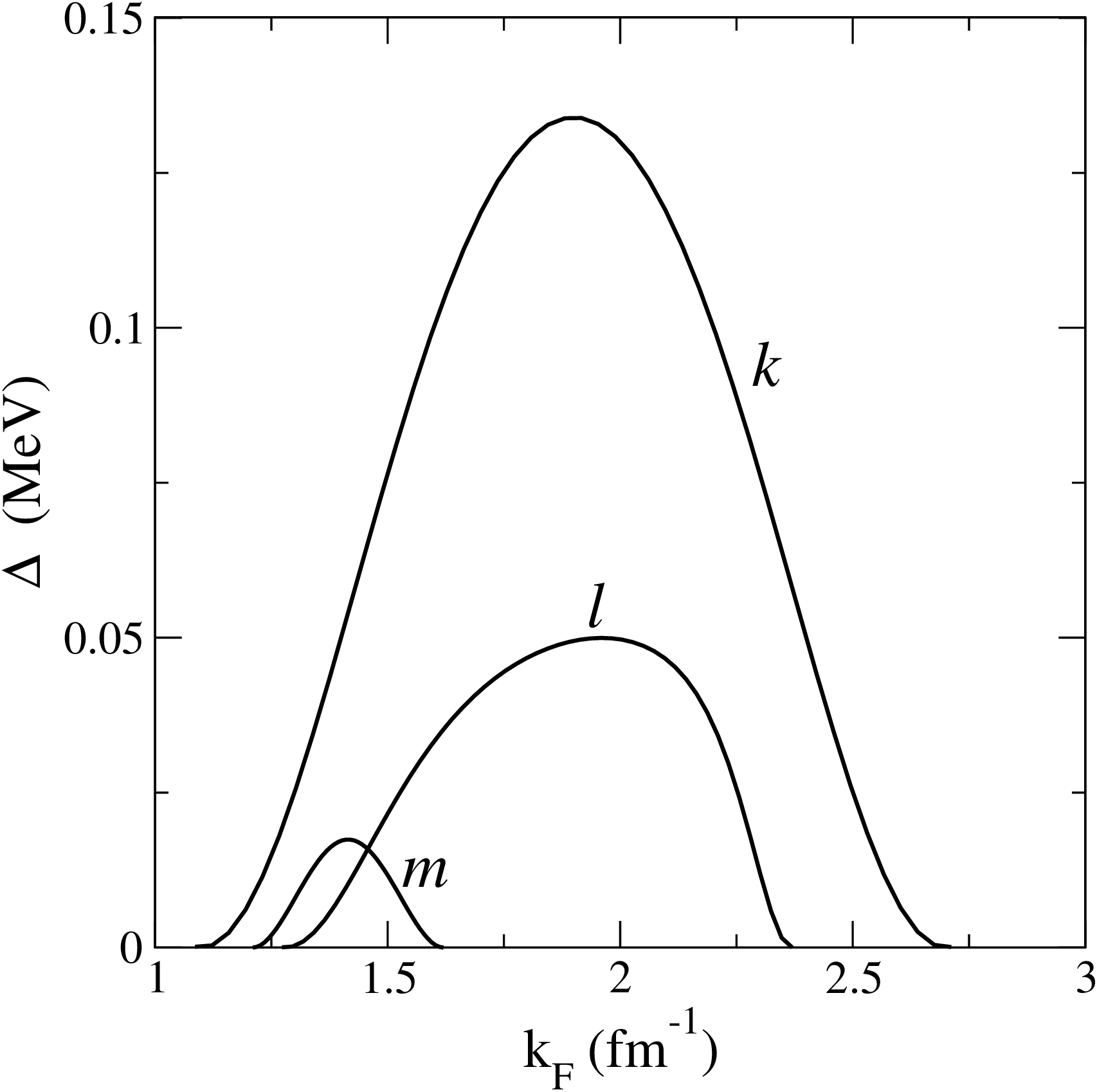}}
\caption{The various energy gaps for the $^3P_2$ neutron pairing, corresponding to the data in Table~\ref{gaptable}, 
are shown as functions of $k_F$. The two panels corresponds to strong and weak neutron superfluidity in 
the core, respectively. }
\label{3p2gaps}
\end{figure}

In Table~\ref{gaptable} we provide data for a set of gap models
based on the literature~\footnote{It should be noted that $\Delta(k_F)$ is only well defined in the isotropic case. For triplet
state pairing, the gap varies over the Fermi surface and one must be careful to choose a ``characteristic'' gap.}.
The models have been selected to span the current range of
uncertainty. The given parameters lead to
good representations of the original results, 
although it should be noted that we have not worried too  
much about the numerical precision of each individual gap. The pairing gaps obtained from 
the data in Table~\ref{gaptable} are shown in Figures~\ref{1s0gaps} and \ref{3p2gaps}.

Having surveyed the current thinking about the nucleon superfluid energy gaps, 
and modelled the results in a useful way, we want to use the results to derive, 
first of all, the associated transition temperatures and then turn our attention 
to the suppression of shear viscosity. As discussed in detail by Yakovlev et al
\cite{yakov} (see also \cite{BCLL90,BCLL92} for useful approximations), the critical temperatures
can be approximated by
\be
{k_B T_c \over \Delta (k_F)} \approx \left\{ \begin{array}{lll} 0.5669 \quad \mbox{ for } ^1S_0 \\ 0.8416 \quad  \mbox{ for } ^3P_2\ (m_J=0) \\
0.4926 \quad \mbox{ for } ^3P_2\ (|m_J|=2) \end{array} \right.
\label{tcr}\fe
Here the effect of the anisotropy of the $^3P_2$ channel, for which $m_J$ represents the 
projection of the angular momentum of the Cooper pair on the quantisation axis, 
is apparent. (Note that, even though the $|m_J|=1$ case has not  been studied
in detail it is expected to be qualitatively similar to the $m_J=0$ case.) 

\begin{figure}
\centerline{\includegraphics[height=8cm,clip] {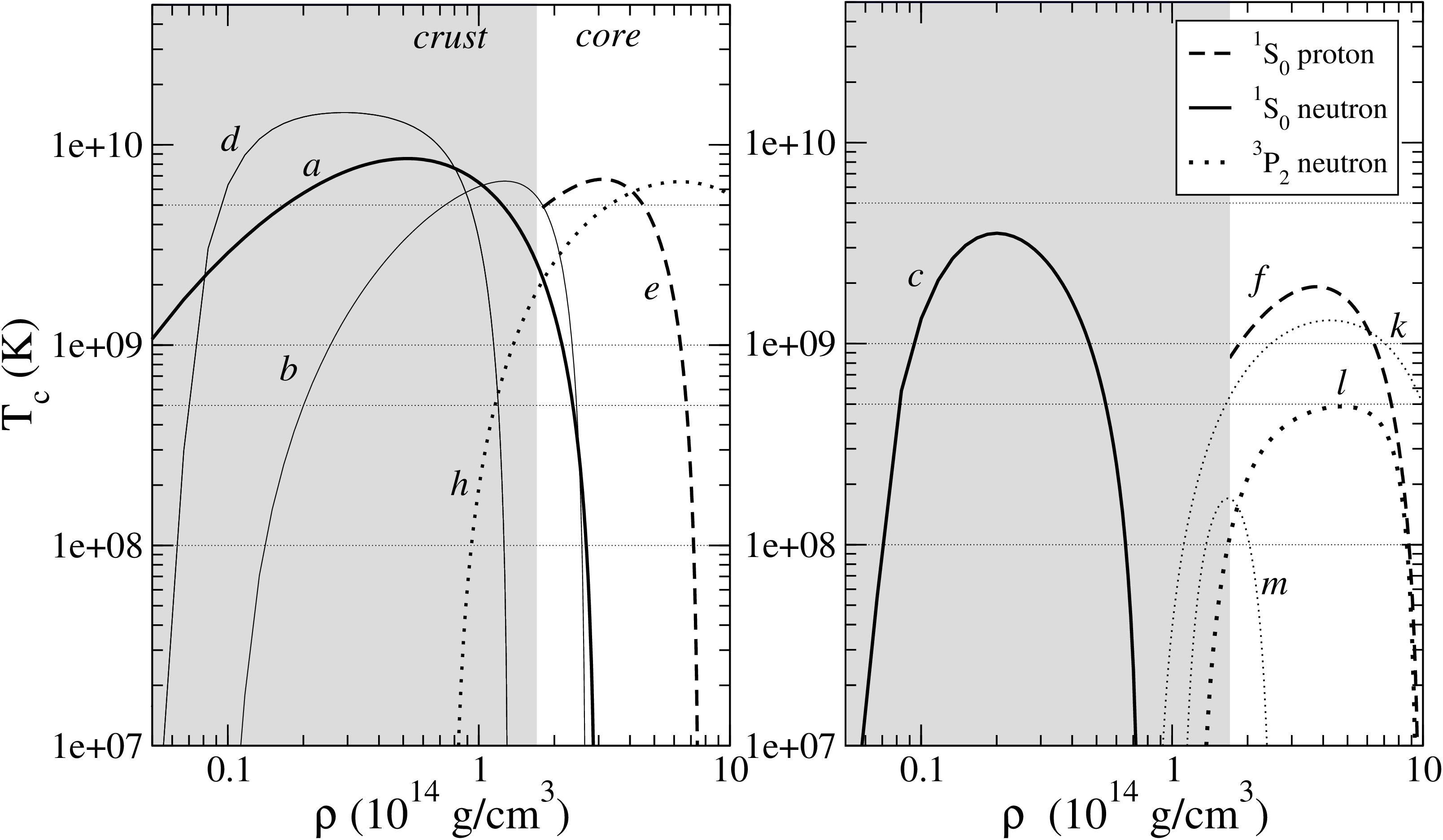} }
\caption{Critical temperatures associated with the various gap models listed in Table~\ref{gaptable}. For clarity we only 
show the $^3P_2$ neutron results for the $m_J=0$ pairing. The results for the $|m_J|=2$ case would be
suppressed by a factor of roughly $0.6$, cf. (\ref{tcr}). As in Figure~\ref{ee-fig} the grey region represents the
neutron star crust, in which there are no free protons and hence no corresponding $T_c$. The inset in the right panel 
helps identify the two sets of gap-models we combine to describe ``strong'' and ``weak'' superfluidity. }
\label{tcrit}\end{figure}

If we combine these approximations with our various gap models (and the data 
for the PAL equation of state from (\ref{palfit})) we obtain the results shown in Figure~\ref{tcrit}.
The two panels distinguish between cases that can  be refered to as representing ``strong''
and ``weak'' superfluidity. We also distinguish the crust and core regions. To make this distinction 
we have assumed that the transition takes place at $0.6$ of the nuclear saturation density
$\rho_0 \approx 2.8\times10^{14} \mbox{ g/cm}^{3}$, i.e. at $\rho_c \approx 1.7\times10^{14} \mbox{ g/cm}^{3}$. 
It should be noted that there are no free protons in the crust, 
and hence no proton gap in that region (we are not accounting for the possibility that nuclei may 
exhibit pairing). It is also worth commenting on the fact that the models
for the $^1S_0$ neutron gap in the crust do not account for interactions with the crust nuclei 
in detail. There are many difficulties associated with this analysis which need to be addressed 
in future research. Anyway, from Figure~\ref{tcrit} we see that superfluid neutrons are expected to be present 
in the crust for temperatures above a few times $10^9$~K.  The same is true for the 
proton superfluid in the core. This means that, all but newly born (less that a month old?) neutron stars
should contain both superfluid neutrons and superconducting protons in the $^1S_0$ phase. However, it is
not clear whether one should expect the corresponding regions in the star to overlap.
Figure~\ref{tcrit} certainly gives examples of $^1S_0$ neutron superfluids which are entirely
confined to the crust. 

At this point it is relevant to mention possible constraints on the 
neutron star superfluidity provided by glitch data. By assuming that the 
glitches are associated with the neutron superfluid in the inner crust,  
one can put constraints on the corresponding moment of inertia required 
to explain the observations. This leads to a requirement that at least 
$1.5-2.5$~\% of the star's moment of inertia is in the superfluid, 
see for example \cite{alp,lel}. Comparing to the models given in 
Table~\ref{gaptable}, this constraint could mean that model $d$ 
(and possibly also model $e$) is ruled out. Of course, this conclusion has many caveats and one should avoid making definitive statements given the lack of 
truly quantitative glitch models.    

As we have already discussed, the results for the $^3P_2$ neutron gap 
are associated with a great deal of uncertainty. From Figure~\ref{tcrit} we see that the 
critical temperature may, in fact, be as low as $10^8$~K. This extreme case would mean that 
the core of an accreting neutron star in a low-mass X-ray binary, for which nuclear burning in 
the crust is expected to equilibrate the temperature around a few times $10^8$~K, may not contain 
superfluid neutrons at all. It should, of course, be emphasised that most of the models
have significantly higher critical temperatures for the $^3P_2$ superfluidity. It would certainly 
be surprising, given the present results, if a neutron star with core temperature 
at the level of $10^8$~K did not contain a region where superconducting protons coexist 
with superfluid neutrons in the $^3P_2$ state.

Next, introducing the variable $\tau = T/T_c$, we learn from Ref.~\cite{yakov} that the 
temperature dependency of the gap functions can be approximated by
\be
y = { \Delta (T) \over k_B T} \approx \left\{ \begin{array}{lll}
\sqrt{1-\tau} \left( 1.456 - { 0.157 /\sqrt{\tau}} + {1.764 / \tau} \right) \quad \mbox{ for } ^1S_0 \\ 
\sqrt{1-\tau} \left( 0.7893 + { 1.188 / \tau} \right) \quad \mbox{ for } ^3P_2\ (m_J=0) \\
{\sqrt{1-\tau^4}}\left( 2.030 - 0.4903\tau^4 + 0.1727 \tau^8 \right)/\tau \quad \mbox{ for } ^3P_2\ (|m_J|=2) \end{array} \right.
\label{tau}\fe
Combining these results with the critical temperatures from (\ref{tcr}) we 
can estimate the pairing gaps at finite temperature. This is a key ingredient 
in constructing the superfluid suppression factor ${\cal R}_p$, 
which represents the extent to which the superconducting protons
no longer scatter the electrons. To estimate this factor we use eq.~(45)
from \cite{gnedin}, which can be written
\begin{eqnarray}
{\cal R}_p &\approx& \left\{ 0.7694 + \sqrt{(0.2306)^2+(0.07207 y)^2 } + (27 y^2 + 0.1476y^4) \exp \left[ 
-\sqrt{(4.273)^2+y^2}
\right] \right. \nonumber \\
 &+& \left. 0.5051 \left[ \exp(4.273-\sqrt{ (4.273)^2+y^2}) - 1 \right] \right\} \exp
\left( 1.187  - \sqrt{(1.187)^2 + y^2}\right) 
\end{eqnarray}
The form of this fit to the numerical results was chosen to lead to  the  limits, ${\cal R}_p \to 1$ as $y\to 0$ and 
 ${\cal R}_p \sim 0.2362 y \exp (-y) $ as $y\to \infty$. For 
temperatures above $T_c$ one should simply use ${\cal R}_p=1$. 
It is perhaps worth pointing out that the exact form of this suppression factor is only relevant
 very close to the transition temperature,  and that an accurate model
may only 
be important for a small subset of neutron star models. 

We need an analogous suppression factor for 
neutron superfluidity. In this case we have carried out a fit to the expressions 
for ${\cal R}_{n1}$ and ${\cal R}_{n2}$ given by eqn. (45) in \cite{baiko}. 
Thus we find 
\be
{\cal R}_n \approx \left[ 0.9543 - \sqrt{ (0.04569)^2 + (0.6971 y)^2} \right]^3 
\exp\left( 0.1148 - \sqrt{ (0.1148)^2+4y^2} \right)
\fe
obviously with $y$ evaluated for the relevant pairing gap according to (\ref{tau}). 
Given this suppression factor, we use
\be
\eta_n \longrightarrow  {\cal R}_n \eta_n
\fe
for the shear viscosity due to neutron-neutron scattering.

\begin{figure}
\centerline{\includegraphics[height=8cm,clip] {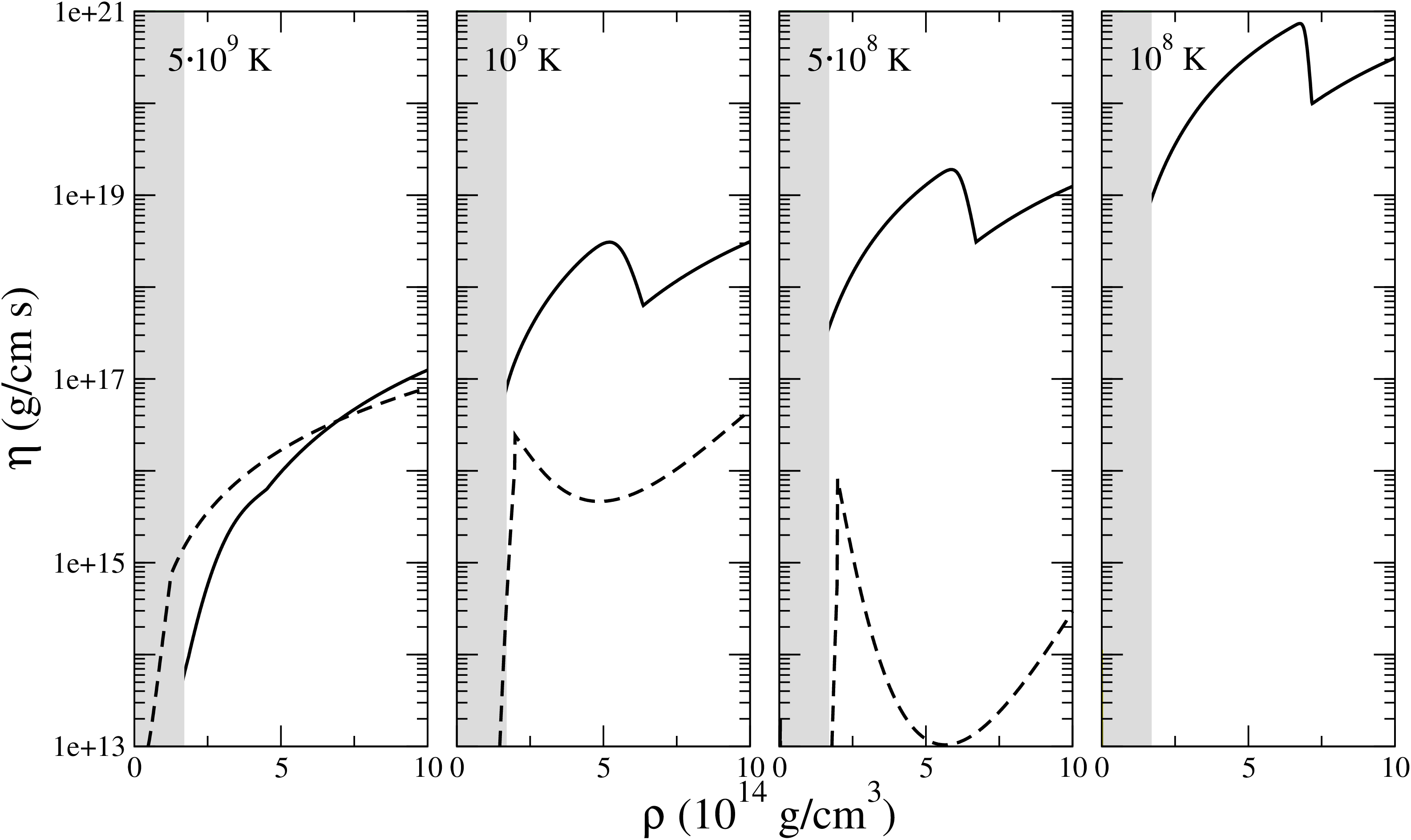} }
\caption{Viscosity coefficients for the  ``strong'' superfluidity case discussed in the text. 
The dashed line represents the neutron-neutron scattering viscosity, while the solid 
line shows the electron viscosity. The grey region shows the extent of the neutron star crust. Note that, in the final panel the neutron viscosity is suppressed to levels below those shown.}
\label{strongvisc}\end{figure}

\begin{figure}
\centerline{\includegraphics[height=8cm,clip] {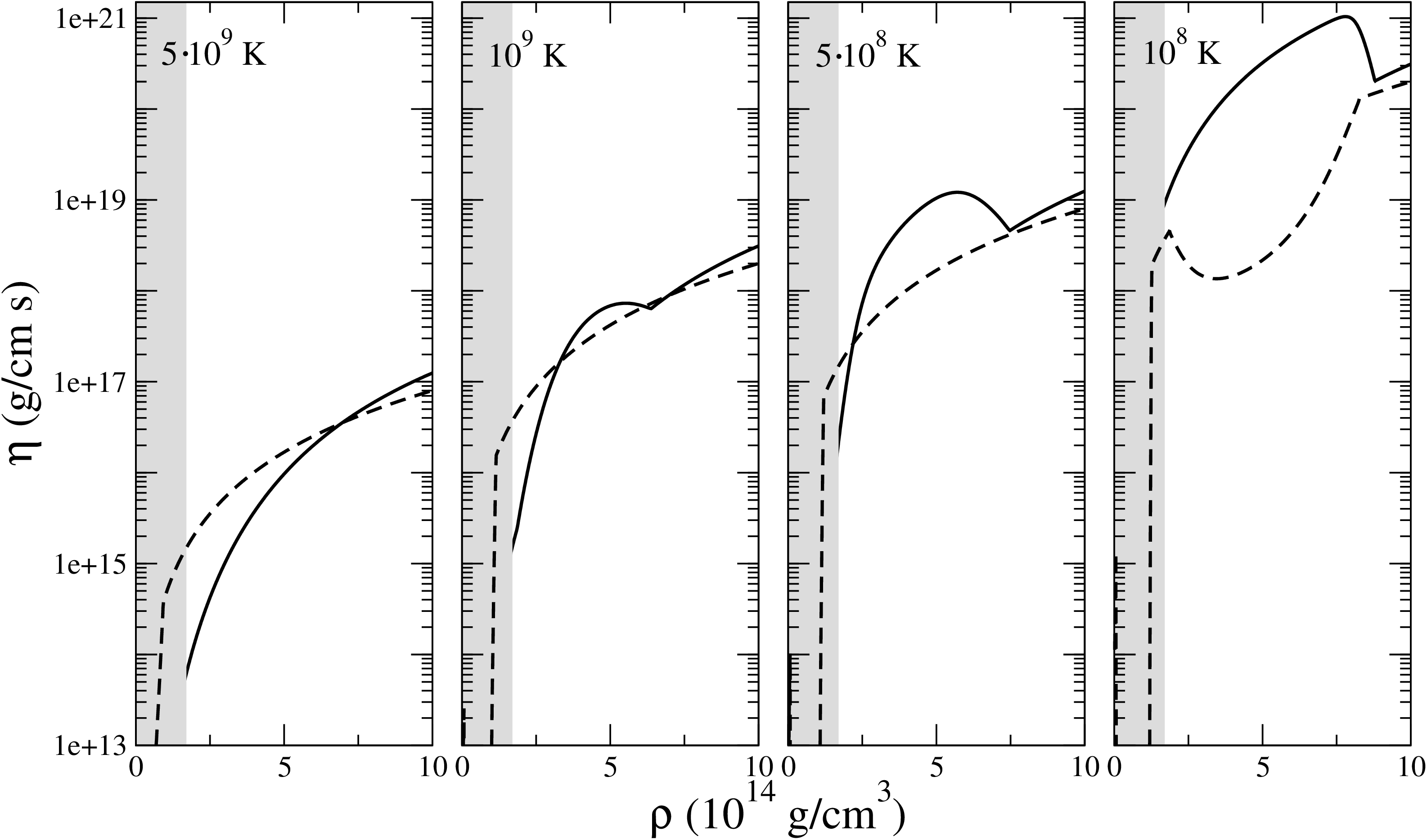} }
\caption{Viscosity coefficients for the  ``weak'' superfluidity case discussed in the text. 
The dashed line represents the neutron-neutron scattering viscosity, while the solid 
line shows the electron viscosity. The grey region shows the extent of the neutron star crust.}
\label{weakvisc}\end{figure}

If we now combine all the results we have discussed, we can calculate 
the electron shear viscosity coefficient as a function of the density 
for various neutron star temperatures. In Figures~\ref{strongvisc} and
\ref{weakvisc} we show the result for strong and weak superfluidity, 
respectively. The former model combines gaps $a$, $e$ and $h$  from Table~\ref{gaptable}, 
while the latter follows if we use gaps  $d$, $f$ and $l$. The two Figures show the 
shear viscosity as a function of the density for four different 
temperatures $T=10^8$~K, $5\times10^8$~K, $10^9$~K and $5\times10^9$~K, also indicated by horisontal 
dotted lines in Figure~\ref{tcrit}. The illustrated data shows (very nicely) 
how the neutron viscosity becomes strongly suppressed at lower temperatures. 
One can also clearly distinguish, eg. from the transition 
that takes place at a density of about $7\times10^{14} \mbox{ g/cm}^3$
in the last three panels of Figure~\ref{strongvisc}, the region where the protons are
superconducting and the (stronger) electron-electron viscosity dominates.  
Taken together, the two figures provide a useful insight into the complexity of the 
neutron star interior and how the viscosity changes as the star matures.

These figures represent the final results of this paper. Having reached the stage where we 
can generate this data, we are now able to turn our attention to the implications for 
neutron star dynamics. In particular, we should be able to study the impact of nucleon 
superfluidity on neutron star oscillations in more detail than has been possible so far. 
We can also return to well-known problems like the ``spin-up'' problem and re-assess
the viscosity due to the core-crust interaction and the associated Ekman layer. 

\section{Future challenges}

Hoping to stimulate further work on the many relevant
issues, we have discussed the effects of superfluidity on
the shear viscosity in a neutron star core. As should be clear
from our analysis, an understanding of this problem requires 
input from several different areas of research, most notably
nuclear physics and relativistic astrophysics. 
Although the results we have collated may 
not be original, we are not aware of any previous study which combines 
the data into a consistent description of the shear viscosity in a 
superfluid neutron/superconducting proton mixture. This is in sharp 
contrast to the many discussions of the role of superfluidity in 
neutron star cooling, see for example \cite{KHY01,YKG01,KYG02,YGKP02,page}. 
We believe that this paper provides 
a useful update on the core viscosities which should be 
valuable for future modelling of neutron star oscillations
and associated instabilities \cite{narev}. 

We have provided a simple model for the electron viscosity
which is relevant both when the protons form a normal fluid and
when they become superconducting. This result explains (as in \cite{easson})
why it is natural that proton superconductivity leads to a 
significant strengthening of the shear viscosity. It also clarifies 
some confusion associated with the description of Cutler and Lindblom
\cite{cl87}. In particular, it is clear that the superfluidity of the 
neutrons is not the key factor which leads to electron-electron
scattering becoming the main  shear viscosity agent. 
Rather, it is the fact that the onset of superconductivity
suppresses the electron-proton scattering. This is an important point.
As can be seen from Figure~\ref{ee-fig}, the electron-electron shear
viscosity is not too different from the result for neutrons scattering off
of each other. This means that, in the temperature range where 
shear viscosity dominates, the  damping 
of neutron star oscillations will be quite similar (modulo multi-fluid effects) in the extreme cases 
when i) the neutrons and protons are both normal, and ii) when the 
neutrons and protons are superfluid/superconducting, respectively.
The contrast with the case when the neutrons are superfluid and the 
protons normal (and viscosity is dominated by $\eta_{ep}$) is clear from Figure~\ref{ee-fig}. This is an 
 interesting observation because it shows that proton 
superconductivity (or rather absence thereof) 
could have a significant effect on the dynamics of a neutron star core. 

Our results are in a form which permits the 
use of data for any given modern equation of state. In contrast to the often used expression obtained 
by Cutler and Lindblom \cite{cl87}, our final formulas are explicitly 
dependent on the proton fraction (as well as the total mass density). 
This is a crucial difference because of the simple fact that the 
result in  \cite{cl87} was obtained for the now seriously 
outdated equation of state derived by Baym, Bethe and Pethick \cite{bbp} in 
1971. We can  obtain an 
indication of the likely ``error bars'' associated with the
electron shear viscosity from
(\ref{eta_ee}) which suggests that $\eta_e \sim x_p^{3/2}$. As 
the proton fraction varies by up to perhaps a factor of two for 
different equations of state we can expect the
viscosity coefficient to be uncertain at the level of (at least) 
a factor of three.

Our discussion also highlights 
a number of  challenges that must be met
if we are to make further progress. First of all, it is clear that an understanding of the 
various superfluid gaps is crucial. We 
believe that the results summarised in Table~\ref{gaptable}, i.e.
the  ``phenomenological'' models we have constructed 
from a range of results in the relevant literature (in line with the 
philosophy adopted in \cite{KHY01,YKG01,KYG02,YGKP02}), provide a useful survey of the 
theoretical range of possibilities. These models are also 
readily included in quantitative studies like that discussed in this paper. 
Of course, many uncertainties remain concerning the superfluid 
parameters. Most important are the suppression factors for the 
various scattering rates which are  needed to evaluate the viscosity coefficients
(${\cal R}_p$ and ${\cal R}_n$ in our analysis). 
Any detailed discussion of these coefficients is certainly valuable. Especially since it 
allows us to make the analysis of temperatures near the
critical temperature 
$T_c$ quantitative rather than qualitative (as in the present discussion).  

In addition to these problems, there are many related challenges.
Although it is commonly acknowledged
 that superfluid components play a crucial role in neutron star dynamics 
(eg. in pulsar glitches \cite{lyne,cheng,alp}), our understanding of the multi-fluid
aspects and the relevant dissipation mechanisms can be improved considerably. 
That this area provides exciting possibilities is nicely
illustrated by the demonstration that a two-stream instability may operate
in superfluid systems with relative flow \cite{twostream}.
Of particular importance for the physics in the neutron star core
will be the superfluid vortices (and magnetic fluxtubes in the case of a 
superconductor), leading to i) dissipation due to 
mutual friction \cite{alpar,men91} 
and ii) potential vortex pinning to the crust
(see, for example, the discussion in \cite{kinney}). 
As the parameters required to model dynamical superfluids (like the 
entrainment coefficients \cite{cj,chamel}) are not yet well 
constrained, this is a research area in its infancy.

Yet another illustration of this fact is provided by the recent discussion
following the likely observation of free precession in PSR B1828-11 \cite{ianj,stairs}.
Of particular relevance for the present paper is Link's 
suggestion \cite{link} that the interaction between neutron vortices and proton 
fluxtubes in a type-II superconductor would lead to a strong 
coupling between the two fluids
and prevent long-period precession. As an alternative, Link 
suggests that the protons could form a type-I superconductor, from which the 
magnetic flux would be expelled. This idea prompted a recent analysis
by Buckley et al \cite{buckley}, who find that the interaction between neutron and proton 
Cooper pairs may indeed lead to the neutron star core becoming a 
type-I superconductor. 
Exactly how this would affect the 
various viscosity estimates, and the dynamics 
of the neutron star core, is  unclear at this point. 
 
Models of the neutron star crust almost exclusively assume that it
can be described as a regular bcc lattice.  At some level this will be an oversimplification. In particular,   it is  likely  
energetically favourable for the  
nuclei to have a variety of ``shapes'', ranging from droplets to rods to plates, as the core-crust 
phase-transition is approached \cite{lorentz}.  
An effective description of these exotic phases
 may  differ significantly from the standard picture.
In addition to this, one should account for the presence of 
superfluid neutrons throughout the bulk of the crust. The 
relevant parameters for this superfluid phase must be considered 
as poorly known since the various gap-calculations do not (yet) 
account for the presence of the nuclei. 
This constitutes a  difficult theory problem, but the result
could be crucial for a number of scenarios ranging from the 
pulsar glitches to gravitational-wave driven mode-instabilities. 
An indication of how different the parameters of the crust 
superfluidity may be compared to those in the core fluid is 
provided by the work of Carter, Chamel and Haensel \cite{cch1,cch2,cch3}.
Their analysis of the entrainment near the base of the crust
shows that the results differ greatly from those typically used to 
model the outer core. The essential difference is in the effective mass
of the neutrons. As in the other cases, the dynamical repercussions
of this result remain to be investigated. 

Finally, let us turn to the physics of the neutron star inner core. 
It is generally expected that, at densities above a few times nuclear 
saturation density, it will be energetically favourable for exotic 
phases of matter to be present. The two most commonly considered cases concern 
hyperons and deconfined quarks. In the first of these cases it has been
pointed out that hyperons provide an efficient refrigerant, since direct URCA 
reactions will be in operation. These reactions would cool 
the star much faster than indicated by the observational 
data. The only way to avoid contradictions with the observations is 
to appeal to hyperon superfluidity  \cite{ppls00}, which would quench the reactions
and slow down the cooling rate. It is also worth pointing out that 
the fast nuclear reactions in a hyperon core are thought to lead
to rapid damping of oscillations due to the resultant bulk viscosity  \cite{jones,lohyp}.
This effect would  be significantly suppressed by superfluidity \cite{haensel}.
In the context of the present paper it is relevant to make two
comments. First of all, it is known that an increase of the hyperon 
population tends to lead to a significant depletion of the electron 
number density. If we suppose that all the other components 
(neutrons, protons and hyperons) are superfluid, then the electrons may provide the only
channel for shear viscosity. If their number density is depressed to 
say $x_e\sim 10^{-3}$ then we would estimate that the shear viscosity is 
something like five orders of magnitude weaker than that illustrated
in Figure~\ref{ee-fig}. The dynamical effects of this could be important.
Secondly, it is worth pointing out that neither the many possible 
vortex-vortex and electron-vortex interactions, nor the multi-fluid 
aspects of superfluid hyperon mixtures have yet been addressed.  
 
The situation is quite similar for the deconfined quark core.
The mixture of s, d, and u quarks is thought to be able to exhibit 
a variety of  ``colour superconducting'' pairings \cite{arw}. Whether this 
leads to multi-fluid dynamics is not clear, although it seems 
likely for at least some of the possible phases. There 
have not yet been many attempts to estimate the relevant 
transport coefficients. Yet, it is clear that 
such studies are of great importance. Especially since they may help 
unveil the true ground state of matter. 
In this context, it is worth mentioning an interesting result. 
Madsen \cite{madsen} has combined 
estimated viscosity coefficients for strange stars with the 
radiation reaction results for the r-mode instability to 
demonstrate that, if the fastest spinning pulsar were 
a pure colour-flavour locked quark star then it ought to be unstable (see \cite{manuel} for 
an alternative). 
One interpretation of this result 
is that this particular form of quark pairing is not present in the 
bulk of a compact star. 
There are of course many caveats to this statement, 
but it is nevertheless relevant. In particular, it  
demonstrates the need to put current and future models of exotic neutron star 
physics
in an astrophysical context and ask whether observations provide useful
constraints on the theory. It is particularly important to consider this 
at the present time when gravitational-wave observations of compact 
object dynamics seem a definite possibility. 

Taking the situation at face value, neutron star (astro)physics 
is a vibrant area of research which 
provides a number of exciting challenges for the future. In order to 
meet these challenges we need a closer 
dialogue between researchers in different 
fields, like nuclear physics and relativistic astrophysics. We hope that 
this paper will help stimulate such discussion.

\section*{Acknowledgements}

We are grateful to Dima Yakovlev for a number of helpful comments on a preprint version of this paper, 
and for sharing unpublished results with us. 
We also thank Mark Alford, Curt Cutler, Michael Gusakov and 
Pawel Haensel for useful discussions.  NA and KG 
are grateful for support from PPARC grant PPA/G/S/2002/00038.

\end{document}